# Uncovering High Thermoelectric Figure of Merit in (Hf,Zr)NiSn Half-Heusler Alloys

L. Chen,[1] S. Gao,[1] X. Zeng,[2] A. Mehdizadeh Dehkordi,[3] T. M. Tritt,[2,3] and S. J. Poon[1,a)]

[1] *Department of Physics, University of Virginia, Charlottesville, Virginia 22904-4714*

[2] *Department of Physics and Astronomy, Clemson University, Clemson, South Carolina 29634-0978*

[3] *Materials Science & Engineering Department, Clemson University, Clemson, South Carolina 29634*

Abstract

Half-Heusler alloys (MgAgSb structure) are promising thermoelectric materials. RNiSn half-Heusler phases (R=Hf, Zr, Ti) are the most studied in view of thermal stability. The highest dimensionless figure of merit (ZT) obtained is ~1 in the temperature range ~450-900$^{\circ}$C, primarily achieved in nanostructured alloys. Through proper annealing, ZT~1.2 has been obtained in a previous ZT~1 n-type (Hf,Zr)NiSn phase without the nanostructure. There is an appreciable increase in power factor, decrease in charge carrier density, and increase in carrier mobility. The findings are attributed to improved structural order. Present approach may be applied to optimize the functional properties of Heusler-type alloys.

[a)] Author to whom correspondence should be addressed; electronic mail: sjp9x@virginia.edu

Half-Heusler alloys (space group F$\bar{4}$3m, C1$_b$) are promising thermoelectric materials due to their high thermopower and moderate electrical resistivity.[1-4] These alloys have composition XYZ, where X and Y denote transition or rare earth elements and Z denotes a main group element. In particular, the RNiSn-type half-Heusler (HH) alloys, where R=Hf, Zr, Ti, are the most investigated to date due to their thermally stable[5] and mass producibility.[6] However, their thermoelectric (TE) figure of merit ZT was limited by the rather high thermal conductivity. ZT is defined as ZT=($S^2\sigma/\kappa$)T, where S, $\sigma$, and $\kappa$ are Seebeck coefficient, electrical conductivity, and thermal conductivity, respectively. $S^2\sigma$ (or $S^2/\rho$) is the power factor (PF), where $\rho$ is electrical resistivity. On the other hand, the study of the effect of intrinsic disorder on TE properties of HH phases has recently gained attention.[7-10] HH phases are prone to antisite disorder, as shown in the inset of Figure 1.[11-13] This intrinsic disorder can adversely influence the PF and thus ZT. To date, ZT of refractory HH phases has remained ~1 at T~700-1100 K, mainly achieved in the nanostructured phases in which significant grain growth appears to be inevitably.[14-23] Although high ZT~1.5 was reported for Sb-doped (Ti,Hf,Zr)NiSn,[24] the results have not been able to be reproduced by other groups. Meanwhile, effort also exists to optimize alloy composition and search for new ones.[21,25] Despite the materials issues mentioned, HH phase based TE modules were found to have a high device efficiencies reaching 20%.[15,26]

From the crystal chemistry viewpoint, antisite disorder seems inevitable.[7,11,12] However, except for long term annealing significantly below the melting point, there is a dearth of effort in directly tackling this basic materials issue. Antisite disorder is believed to persist even after the annealing. The electronic structure that determines the band mass and carrier mobility are sensitive to lattice disorder. In this paper, we report an approach to improve the ZT of n-type (Hf$_{0.6}$Zr$_{0.4}$)NiSn$_{0.995}$Sb$_{0.005}$. This has led to enhanced power factor that raises ZT to 1.2, overcoming the apparent ZT~1 barrier without the need of *nanostructures* or *composition optimization*. Tantalizing correlations between structural order and thermoelectric properties are observed. Our findings show that the intrinsic disorder that has plagued the TE performance of half-Heuslers can be reduced if the synthesis and annealing processes are carried out near the melting temperature. Similar materials issues may exist in other alloy systems. Thus, the present approach may be applied in order to improve the thermoelectric properties of other materials.

$Hf_{0.6}Zr_{0.4}NiSn_{0.995}Sb_{0.005}$ with ZT~1 obtained in spark plasma sintered samples[15] is selected for investigation. In order to focus on the issue of intrinsic disorder, the influence due to extrinsic effects must be minimized at the outset. As such, Ti-containing alloys were deliberately excluded in order to avoid phase separation,[27] which could mask the intrinsic structural effect being investigated. In comparison, Hf and Zr have the same atomic size.[28,29] The precursor materials for SPS consisted of pulverized ingots. Ball milling was not employed. Ingots were prepared by arc melting under an Argon atmosphere. Ingots were pulverized into 10 to 50 μm size powders followed by consolidation using Spark Plasma Sintering (Thermal Technologies SPS 10-4) technique. The sintering temperature needed to produce a fully dense compact was around 1100°C. Samples were sintered at 1100°C, 1250°C, and 1350°C for a few minutes under 60 MPa followed by annealing at the same temperatures for 30 minutes. Investigation of microstructure and composition was performed by the combined use of PANalyticalX'Pert Pro MPD instrument and FEI Quanta 650 Scanning Electron Microscope. Resistivity and thermopower measurements were performed using ZEM3 system. Thermal conductivity was calculated as the product of the specific heat (Netzsch Differential Scanning Calorimeter), the thermal diffusivity (Netzsch LFA 457 MicroFlash system) and the mass density. The lattice thermal conductivity $\kappa_L = \kappa - \kappa_e$ was obtained by knowing $\kappa_e$, estimated by using the Wiedenann-Franz relationship $\kappa_e = L\sigma T$, where L is the Lorenz Number. The Hall coefficient ($R_H$) was measured in a 1 T field using a Quantum Design physical property measurement system. X-Ray Diffraction (XRD) of the samples reveal single HH phase. The compositional homogeneity checked by EDS map scan over a 90000 μm$^2$ area shows no evidence for any compositional variation across a few μm$^2$.

The as cast ingots contained HH phase as well as about 10-15% other phases Systematic treatment of the ingots was performed to obtain single-phase samples for characterization. Three groups of samples were prepared and characterized, as summarized below:

(i) An ingot was annealed at 700°C for 8 days and another ingot at 1000°C for 2 days, significantly below the melting temperature, $T_{melt}$ ~1460°C,[5] to obtain single phase samples.[1-4,30] X-ray diffraction pattern for the annealed samples is shown in Figure 1. The need for long term annealing indicates large barriers for long-range atomic diffusion. The degree of structural disorder is encoded in the width of the X-ray diffraction peak. Details on peak width analysis are provided in Supplementary Material.[31] Considerable lattice strain remains in the samples even after extended annealing. In addition to anitsite disorder, compositional fluctuation can still exist in the single-phase samples, not detectable by microprobe analysis. Compositional fluctuation will give rise to lattice strain since Hf/Zr and Sn exhibit different atomic sizes in both the metallic and covalent states.[28,29]

(ii) Micron size particulates pulverized from the ingots were consolidated using spark plasma sintering and annealed at various temperatures up to 1350$^o$C, about 100$^o$C below the melting point. At high temperatures, atomic diffusion is accelerated and the composition is more homogeneous as evident by the significant reduction in the lattice strain. Entropy driven *antisite disorder*, however, may still exists. Annealing at still higher temperature, 1400$^o$C, resulted in partial melting and the appearance of about 10% secondary phases. Some strain reduction in the HH phase was detected.

(iii) In an attempt to further reduce *antisite disorder* in the samples studied in (ii), additional annealing was performed at lower temperatures on two samples, one at 900$^o$C for 2 days and the other at 700$^o$C for 8 days. This attempt has resulted in only a very small reduction in the lattice strain.

To summarize the results, lattice strain is plotted as a function of annealing temperature in Figure 2. From the study of microstructure as described in Supplementary Materials,[31] the lattice strain does not change even as the grain size increases from submicron to several micron. For comparison, the annealed ingots described in (i) also comprise micron size grains. Thus, the measured lattice strain is independent of the grain size and only depends on the annealing temperature. The degree of structural disorder is essentially encoded in the observed lattice strain.

To explore the impact of annealing temperature on thermoelectric properties, $Hf_{0.6}Zr_{0.4}NiSn_{0.995}Sb_{0.005}$ samples annealed at 1350$^o$C, 1250$^o$C, 1100$^o$C, 1000$^o$C and 700$^o$C were examined. Figures 3a and 3b show that as the annealing temperature increases, the Seebeck coefficient increases over the entire temperature range accompanied by a decrease in the electrical resistivity, resulting in a significant increase in the power factor. In fact, the power factor increases by as much as 50% as the annealing temperature increases from 700°C to 1350°C, as shown in Figure 3c. Subsequent annealing of the 1350°C-sample at lower temperatures resulted in negligibly small changes in S and ρ, despite the significant growth in grain size mentioned. The large increase in power factor discovered is unprecedented for a monolithic half-Heusler alloy. Carrier trapping, which tends to enhance the thermopower, is ineffective in the absence of nanostructure. Since the beneficial effect to the electronic properties is determined by annealing the samples, structural ordering must have played a key role in the enhancement of PF.

Given the large enhancement of power factor observed, the quest for high ZT is focused on the SPSed samples. These same samples show much lower porosity and are much less likely to exhibit cracking than as cast samples. As such, the thermal conductivity of SPS samples can be readily measured by using the laser flash technique. Thermal conductivity results are shown in Figure 4a and 4b. For the SPSed samples, the total thermal conductivity and lattice thermal conductivity change little for different annealing temperatures, only by ~3% and 5%, respectively. Further, the variation is not systematic. The results also show the usual upturns at high temperatures due to bipolar contribution. ZT is shown as a function of temperature in Figure 4c. The increase in ZT, from 1 to 1.2 for the SPSed samples, is mostly due to increase in the power factor.

Hall effect measurements were conducted in order to help understand the electronic origin of the enhanced power factor. The Hall coefficient ($R_H = -1/nq$) reveals both the carrier type and carrier density, where n is the carrier concentration and q the carrier charge. The carrier mobility ($\mu_H$) is then obtained from the electrical conductivity $\sigma = nq\mu_H$. Previous investigation showed that the electronic transport properties of n-type (Hf,Zr)NiSn half-Heusler alloys could be described by a one-band model.[32] The n and $\mu_H$ values are shown in Table I. It shows that as annealing temperature increases, n decreases and $\mu_H$ increases. Applying the Mott formula to a degenerate semimetal at temperatures below the Fermi temperature,[33] the thermopower is given by the scaling relation $S \sim qm^*T/n^{2/3}$, where m* is the effective band mass. The case for degenerate semimetal is validated by noting that the carrier concentration n is of the same order as or larger than the quantum concentration $n_Q$, where $n_Q = (2\pi m^* k_B T/h^2)^{3/2}$, estimated to be ~$10^{19}$ cm$^{-3}$ at 300K. The relation shows that S increases as n decreases, as observed. The decrease in n, by as much as a factor of near three across the samples studied, is correlated with the decrease in the lattice strain. Correspondingly, the carrier mobility $\mu_H$ increases by nearly a factor of three. As the structural order improves, it also *"strengthens"* the semiconducting state of the half-Heusler phase. Previous studies have alluded to certain imperfections inside the bandgap manifested in the presence of *in-gap* states and thus a reduced bandgap size.[7,34,35] The bandgap can be obtained by using the expression $E_g = 2e|S_{max}|T_{max}$ given by Goldsmid and Sharp, where $E_g$ is the bandgap, $|S_{max}|$ is the magnitude of the maximum thermopower and $T_{max}$ is the temperature at which $|S_{max}|$ occurs.[36] As the annealing temperature increases, a systematic enhancement in the bandgap is observed, as shown in Table 1. The bandgap for these alloys has been estimated to be in the range 0.18-0.50eV.[34,37]

The relationship between carrier scattering and mobility is linked by using the Matthiessen's rule: $\frac{1}{\mu_H} = \frac{1}{\mu_{impurities}} + \frac{1}{\mu_{phonons}} + \frac{1}{\mu_{defects}}$, where $\mu_H$ is the Hall mobility, $\mu_{impurities}$ is the mobility that the material would have due to scattering by ionized donor atoms, $\mu_{phonons}$ is due to phonon scattering and $\mu_{defects}$ is due to lattice defect scattering. According to Brooks-Herring model, $1/\mu_{impurities}$ is proportional to the carrier concentration n.[38] The relationship between carrier mobility and carrier concentration is shown in Figures 5. The dotted line with zero intercept represents an ideal case where there is no lattice defect or phonon scattering. For real material especially alloys, carrier scattering by lattice defects and phonons are significant, and thus the solid line in the figure has positive intercept. The carrier mobility from phonon and defect scattering are dependent on annealing. If the mobility from phonon and defect scattering were independent of annealing, all data would fall on the solid line. However, as the annealing temperature rises the inverse Hall mobility falls increasingly below the solid line. This indicates an increase in mobility due to decreased phonon and defect scattering. The large increase in $\mu_H$ underscores the improvement of structural order. As fluctuations in the periodic atomic potential due to disorder are reduced, then the scattering of carriers is also pronouncedly reduced. This is accomplished predominately by the straightforward synthesis process as described herein without the need for the *addition of nanoparticles, creation of in-situ nanostructures or composition optimization.*

The power factor PF=$S^2\sigma$ scales as $\mu_H/n^{1/3}$. The role of carrier mobility is apparent from the measurement since the increase in $S^2$ (~30%) alone as the lattice strain relaxes is insufficient to account for the larger increase in PF (~50%). The relaxation in the lattice strain and improvement of structural order are expected to have effect on the lattice thermal conductivity $\kappa_L$ shown in Figure 4b. Nevertheless, $\kappa_L$ changes little as the annealing temperature increases. It is cautioned that due to the different grain sizes in the samples, further study is warranted. Future studies will need to consider the optimization of total thermal conductivity.

In summary, thermoelectric properties of n-type (Hf,Zr)NiSn half-Heusler alloys have been investigated by annealing the materials near the melting point in order to reduce the structural disorder, as evidenced by the reduction in the lattice strain. It has led to a significant enhancement of the power factor, by as much as 50%, which raised ZT to 1.2 without the need of *nanostructures*. The large increase in power factor was attributed to the significant decrease of charge carrier density and increase of carrier mobility consistent with the strengthening of the bandgap of the half-Heusler phase. The present approach, if applied to the broader class of Heusler alloys, may result in the improvement of their multifunctional properties.

TABLE I. Hall coefficient ($R_H$), carrier concentration (n), Hall mobility ($\mu_H$) and and Band gap ($E_g$) of $Hf_{0.6}Zr_{0.4}NiSn_{0.995}Sb_{0.005}$ under different synthesis and annealing conditions. Samples prepared by SPS were annealed for 30 minutes at the temperatures shown.

| Condition | Hall coefficient $R_H$ ($10^{-1}cm^3/C$) | Carrier concentration n ($10^{19}cm^{-3}$) | Hall mobility $\mu_H$ ($cm^2/(V*s)$) | Band gap $E_g$ (eV) |
|---|---|---|---|---|
| Anneal 700°C, 8 days | -2.21 | 2.83 | 15.1 | 0.276 |
| Anneal 1000°C, 2 days | -3.45 | 1.81 | 24.0 | 0.284 |
| SPS:1150°C | -3.47 | 1.80 | 27.1 | 0.286 |
| SPS:1250°C | -3.57 | 1.75 | 28.3 | 0.298 |
| SPS:1350°C | -5.17 | 1.21 | 38.2 | 0.317 |
| SPS:1350°C anneal 700°C, 8 days | -5.39 | 1.16 | 38.3 | 0.316 |
| SPS:1350°C anneal 900°C, 2 days | -5.43 | 1.15 | 40.1 | 0.319 |

List of Figures

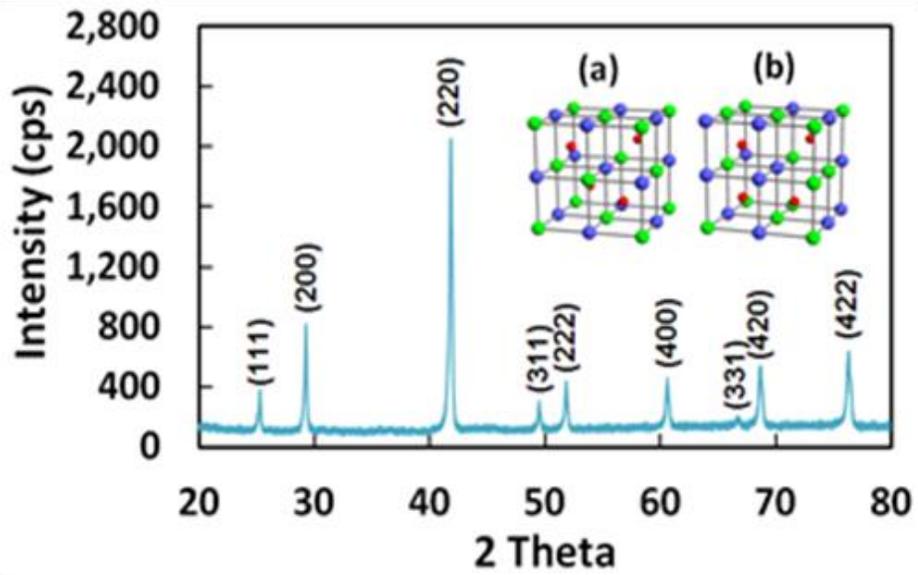

FIG. 1. X-ray patterns for n-type $Hf_{0.6}Zr_{0.4}NiSn_{0.995}Sb_{0.005}$ ingot-based SPSed sample annealed at 700°C for 8 days. All samples annealed up to 1350°C show single-phase diffraction pattern. Inset shows Half-Heusler unit cells with Hf, Zr (green), Sn (blue), and Ni (red) occupying three face-centered-cubic sublattices. Left: the structure is perfectly ordered. Right: *antisite disorder* occurs between Sn and (Hf, Zr) atoms each of which have 4 atoms in the unit cell. Ni can occupy the supposedly vacant fourth sublattice.

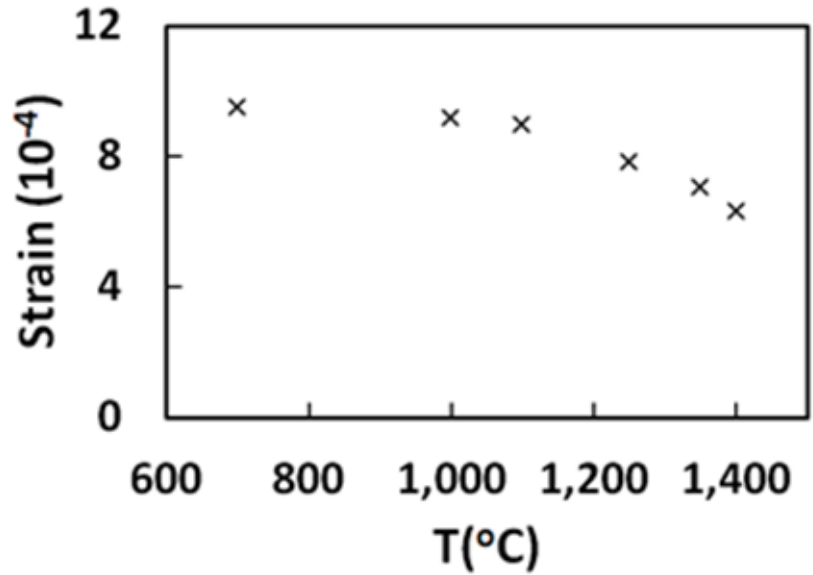

FIG. 2. Lattice strain as a function of annealing temperatures for n-type $Hf_{0.6}Zr_{0.4}NiSn_{0.995}Sb_{0.005}$. Samples annealed at 1100°C and above are those prepared by using spark plasma sintering. Strain values for two samples annealed at 1350°C followed by additional annealing at 900°C and 700°C are similar to that for the sample annealed at only 1350°C. Experimental uncertainty is essentially represented by the size of the data points shown.

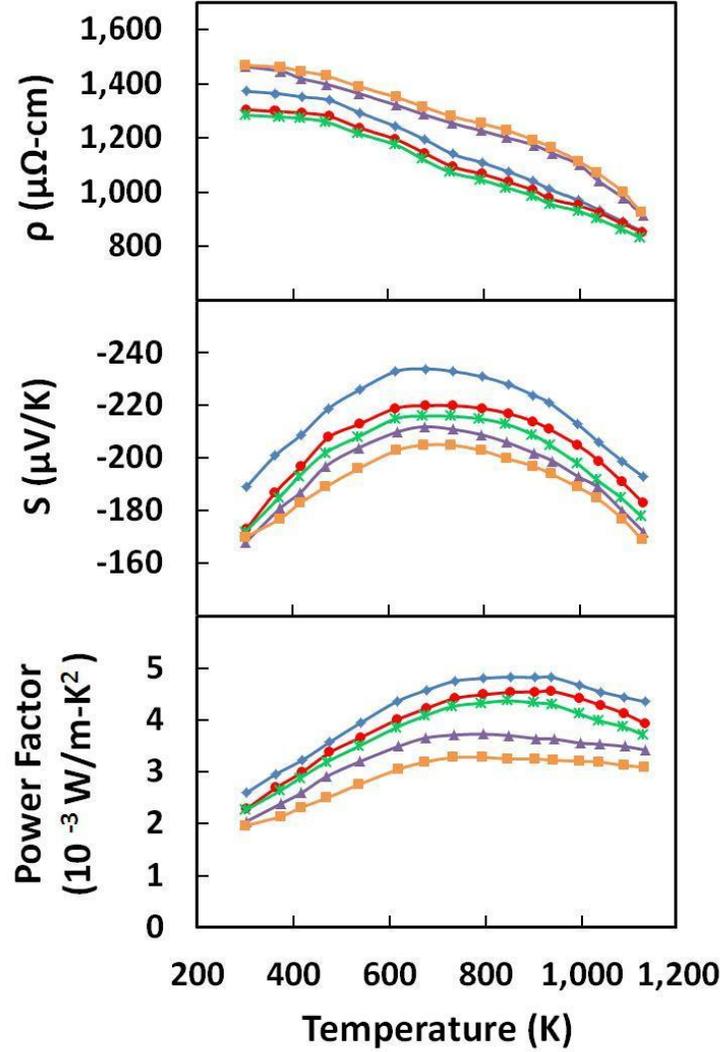

FIG. 3. Thermoelectric properties of n-type $Hf_{0.6}Zr_{0.4}NiSn_{0.995}Sb_{0.005}$ annealed at 1350°C for 30 minutes (blue rhombus), 1250°C for 30 minutes (red circle), 1100°C for 30 minutes (green star), 1000°C for 2 days (purple triangle), 700°C for 8 days (orange square): (a) Electrical resistivity ($\rho$), (b) Seebeck coefficient (S), and (c) Power Factor. Samples annealed at 1100°C and above are those produced by the SPS method.

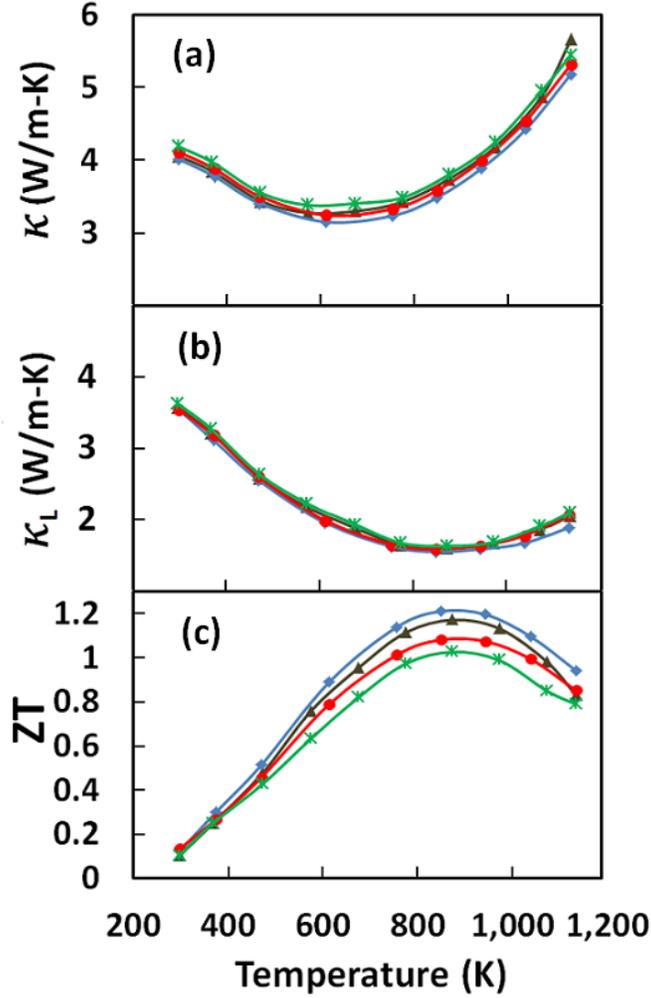

FIG. 4. (a) Thermal conductivity ($\kappa$), (b) lattice thermal conductivity ($\kappa_L$), and (c) ZT of n-type Hf$_{0.6}$Zr$_{0.4}$NiSn$_{0.995}$Sb$_{0.005}$ annealed at 1350°C for 30 minutes (blue rhombus), 1250°C for 30 minutes (red circle), 1100°C for 30 minutes (green star), 1350°C for 30 minutes followed by 700°C for 8 days (dark brown triangle). Given the uncertainty in all the measurements, the resulting ZT has an uncertainty of $\approx \pm 10\%$, which is comparable or less than most other groups.

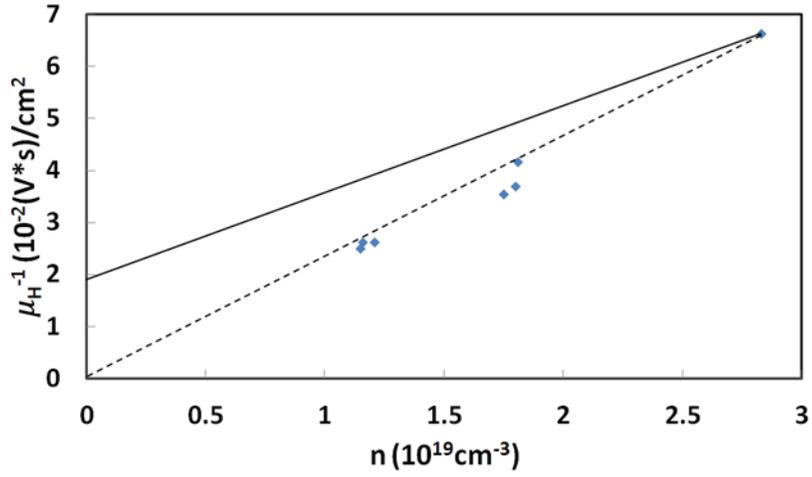

FIG. 5. Inverse mobility as a function of carrier concentration for n-type $Hf_{0.6}Zr_{0.4}NiSn_{0.995}Sb_{0.005}$. Relationship between mobility and carrier concentration by using the Matthiessen's rule with (solid line) and without (dotted line) phonon and defect scattering for the sample annealing at 700 °C.